\documentclass[aps,prl,twocolumn,superscriptaddress,showpacs]{revtex4}
\usepackage{dcolumn}
\usepackage{epsfig}
\def\met{\mathbin{E\mkern - 11mu/_T}}
\begin{document}
\title{\boldmath 
\vspace*{-17mm}
\hfill {\rm CDF/PUB/TOP/PUBLIC/5590}\\[2mm]
\hfill {\rm FERMILAB-PUB-01/318-E}\\[5mm]
Search for Single-Top-Quark Production in $p\bar{p}$ Collisions 
at $\sqrt{s} = 1.8$ TeV}
\date{\today}
\author{D.~Acosta}
\affiliation{University of Florida, Gainesville, Florida  32611}
\author{T.~Affolder}
\affiliation{Ernest Orlando Lawrence Berkeley National Laboratory,  Berkeley, 
California 94720}
\author{H.~Akimoto}
\affiliation{Waseda University, Tokyo 169, Japan}
\author{M.~G.~Albrow}
\affiliation{Fermi National Accelerator Laboratory, Batavia, Illinois  60510}
\author{P.~Amaral}
\affiliation{Enrico Fermi Institute, University of Chicago, Chicago,  Illinois 
60637}
\author{D.~Ambrose}
\affiliation{University of Pennsylvania, Philadelphia,  Pennsylvania 19104}
\author{D.~Amidei}
\affiliation{University of Michigan, Ann Arbor, Michigan 48109}
\author{K.~Anikeev}
\affiliation{Massachusetts Institute of Technology, Cambridge, Massachusetts  
02139}
\author{J.~Antos}
\affiliation{Institute of Physics, Academia Sinica, Taipei, Taiwan 11529,  
Republic of China}
\author{G.~Apollinari}
\affiliation{Fermi National Accelerator Laboratory, Batavia, Illinois  60510}
% ...
\collaboration{CDF Collaboration}
\noaffiliation

\begin{abstract}
We search for standard model single-top-quark production in 
the $W$-gluon fusion and $W^{\star}$ channels using $106$ pb$^{-1}$ 
of data from $p\bar{p}$ collisions at $\sqrt{s} = 1.8$ TeV collected 
with the Collider Detector at Fermilab.  We set an upper limit at 
95\% C.L. on the combined $W$-gluon fusion and $W^{\star}$
single-top cross section of 14 pb, roughly six 
times larger than the standard model prediction.  Separate 95\% C.L.
upper limits in the $W$-gluon fusion and $W^{\star}$ channels are also 
determined and are found to be 13 and 18 pb, respectively.
\end{abstract}
\maketitle

The observation of the top quark in $p\bar{p}$ collisions 
at the Fermilab Tevatron has relied on pair production through the strong 
interaction, typically $q\bar{q}\rightarrow t\bar{t}$. A
top quark can also be produced singly, in association with a $b$ quark, 
through the electroweak interaction \cite{will86}. 
The two dominant ``single-top'' processes are
``$Wg$''(i.e. $W$-gluon fusion, $qg \rightarrow t\bar{b}q^{\prime}$) 
and ``$W^{\star}$'' ($q\bar{q}^{\prime} \rightarrow t\bar{b}$).
Within the context of the standard model, a measurement of the rate
of these processes at a hadron collider allows a determination of
the Cabibbo-Kobayashi-Maskawa matrix element $V_{tb}$ \cite{stel95}.
Assuming $|V_{tb}|=1$, the predicted cross sections for 
$Wg$ and $W^{\star}$ are 1.7 pb \cite{stel98} and 
0.7 pb \cite{smit96} respectively, compared to 5.1 pb for $t\bar{t}$ pair 
production \cite{bonc98}.  The D\O\ Collaboration has 
recently published 95\% confidence level (C.L.) upper limits of 22 pb on 
$Wg$ and 17 pb on $W^{\star}$ production \cite{abaz01}.
In this Letter we report on two searches, one for the two single-top processes
combined, and the other for each process separately.

The expected final state of a single-top event consists of $W$-decay 
products plus two or more jets, including one $b$-quark jet from the decay
of the top quark. In $W^{\star}$ events, we expect a second $b$-quark jet 
from the $W^{\star}t\bar{b}$ vertex.  In $Wg$ events, we expect a jet 
originating from the  recoiling light quark and a second $b$-quark jet produced 
through the splitting of the initial-state gluon. This $b$-quark jet is produced 
at larger absolute value of pseudorapidity \cite{CDF-coordinates} and 
lower transverse momentum than in $W^{\star}$ events \cite{will86}.

Single-top processes are harder to observe than $t\bar{t}$ production because their
cross section is smaller and their final state, containing fewer jets,
competes with a larger $W$+multijet background from QCD.
A priori we do not expect sensitivity to the standard model cross section
in the presently available data. However, a number of new physics processes 
could enhance the single-top production rate, motivating a 
search~\cite{lari97,rosn90}.

Our measurement uses  $106\pm 4$ pb$^{-1}$ of data from $p\bar{p}$ collisions 
at $\sqrt{s} = 1.8$ TeV collected with the Collider Detector at Fermilab
between 1992 and 1995 (``Run I'').  The detector is
described in detail elsewhere \cite{abe88}.
We restrict our single-top search to events with evidence of a leptonic 
$W$-decay:  an isolated \cite{isolation} electron (muon) candidate with 
$E_T$ ($P_T$) $>20$ GeV (GeV/$c$) and missing transverse energy \cite{missing-ET} 
$\met$  $> 20$ GeV from the neutrino.  We remove events that were identified 
in a previous CDF analysis \cite{abe98} as $t\bar{t}$ dilepton 
candidates.  Events with a second, same-flavor and opposite-charge lepton that 
forms an invariant mass with the first lepton between 75 and 105 GeV/$c^{2}$
are rejected as likely to have come from $Z^{0}$ boson decays.  
Furthermore, to reject those dilepton $t\bar{t}$ or $Z^{0}$ candidates where one 
lepton fails our electron or muon identification, we also remove events that 
contain a track with $P_T > 15$ GeV/$c$ and charge opposite that of the primary 
lepton, and such that the total $P_{T}$ of all tracks in a cone of radius 
$\Delta R\equiv\sqrt{\Delta\eta^{2}+\Delta\phi^{2}} = 0.4$ around this track
is less than 2 GeV/$c$ \cite{abe97}.  Jets are formed as clusters of 
calorimeter towers within cones of fixed radius $\Delta R = 0.4$.  Events are 
required to have one, two, or three jets with $E_T >15$ GeV and $|\eta|<2.0$; 
at least one jet must be identified as likely to contain a $b$ quark 
(``$b$-tagged'') using displaced-vertex information from the silicon vertex 
detector (SVX) \cite{abe97}.  If a  second jet in the event is also 
$b$-tagged, either in the SVX or by the presence of a soft lepton indicative 
of semileptonic $b$ decay, the event is labeled ``double-tag'', otherwise it 
is labeled ``single-tag''.  The above event selection cuts are common to our 
combined and separate searches for the two single-top processes.  Additional 
cuts are applied within each analysis.

We first describe our search for single-top production in the $Wg$ and
$W^{\star}$ channels combined.  The expected signal significance 
is improved by requiring the invariant mass $M_{\ell\nu b}$ reconstructed 
from the lepton, neutrino, and highest-$E_{T}$ $b$-tagged jet, to lie in a 
window around the top quark mass, $140 < M_{\ell\nu b} < 210$ GeV/$c^{2}$.  
The neutrino momentum is obtained from the $\met$ and the
constraint that  $M_{\ell\nu}$ = $M_W$ \cite{wmass}.
The variable $M_{\ell\nu b}$ discriminates against both non-top and  
$t\bar{t}$ backgrounds, in the latter case because combinatorial
errors in assigning partons to final-state jets broaden the
$M_{\ell\nu b}$ distribution compared to single top. 

We determine the efficiency of our selection criteria from events
generated by the {\sc pythia} Monte Carlo program \cite{sjos94} and 
subjected to a CDF detector simulation.  The acceptance times branching 
ratio is $(1.7\pm 0.3)\%$ for each of the two single-top processes.
The largest contributions to the acceptance uncertainties
come from lepton triggering and identification (10\%), and $b$-tagging
(10\%).  Combining these acceptances with the cross sections predicted by 
theory \cite{stel98,smit96} and the size of the CDF Run I dataset, we expect 
a total signal yield of $4.3$ events.

Expectations for signal and background rates are listed in the second 
column of Table \ref{tab:pred}.  We estimate the $t\bar{t}$ background 
from a {\sc herwig} Monte Carlo calculation \cite{marc88} followed by 
a detector simulation.  Normalizing to the theoretically-predicted cross 
section, $\sigma_{t\bar{t}} = 5.1\pm 0.9$ pb \cite{bonc98}, 
we expect $8.4\pm 2.7$ $t\bar{t}$ events to survive our selection 
criteria, where the uncertainty includes theoretical and acceptance 
contributions.
\begin{table}
\begin{ruledtabular}
\begin{tabular}{lccc}
        &       Combined Search    & \multicolumn{2}{c}{Separate Search} \\
        &          $W+1,2,3$ jets  & \multicolumn{2}{c}{ $W+2$ jets }    \\
Process &                          &    Single-tag    &   Double-tag     \\
\colrule
$Wg$               & $3.0\pm 0.6$  & $1.4\pm 0.3$     & $0.04\pm 0.01$   \\
$W^{\star}$        & $1.3\pm 0.2$  & $0.55\pm 0.15$   & $0.32\pm 0.06$   \\
$t\bar{t}$         & $8.4\pm 2.7$  & $1.4\pm 0.5$     & $0.7 \pm 0.2 $   \\
non-top            & $54\pm 12$    & $10\pm 2 $       & $1.6 \pm 0.4 $   \\
\colrule
Total              & $67\pm 12$    & $14\pm 2 $       & $2.7 \pm 0.5 $   \\
\colrule
Observed           &      65       &       15         &       6          \\
\end{tabular}
\end{ruledtabular}
\caption{\label{tab:pred}
Expected numbers of signal and background events passing all cuts in the 
$W$+jets data sample, compared with observations.  The uncertainties on 
the expected numbers of single-top events do not include uncertainties 
on the theoretical cross section calculations.}
\end{table}

The largest component of the non-$t\bar{t}$ background in the 
SVX-tagged $W$+jets sample is inclusive $W$ production in 
association with heavy-flavor jets (e.g. $p\bar{p}\rightarrow Wg$, 
followed by $g\rightarrow b\bar{b}$).  Additional sources
include ``mistags,'' in which a light-quark jet is erroneously
identified as heavy flavor, ``non-$W$'' (e.g. direct $b\bar{b}$
production), and smaller contributions from $WW$, $WZ$, and
$Z+$heavy-flavor \cite{abe97}.  
The mistag and non-$W$ rates are estimated 
from data, the $W$+heavy-flavor rates from Monte Carlo normalized 
to data, and the smaller sources such as diboson production 
from Monte Carlo normalized to theory predictions \cite{abe97}.  
The total non-top background expectation is $54\pm 12$ events. The
uncertainty on our background includes the effect of varying the
top mass by its uncertainty of $\pm 5$ GeV/$c^2$.

\begin{figure}
\centerline{
\epsfig{file=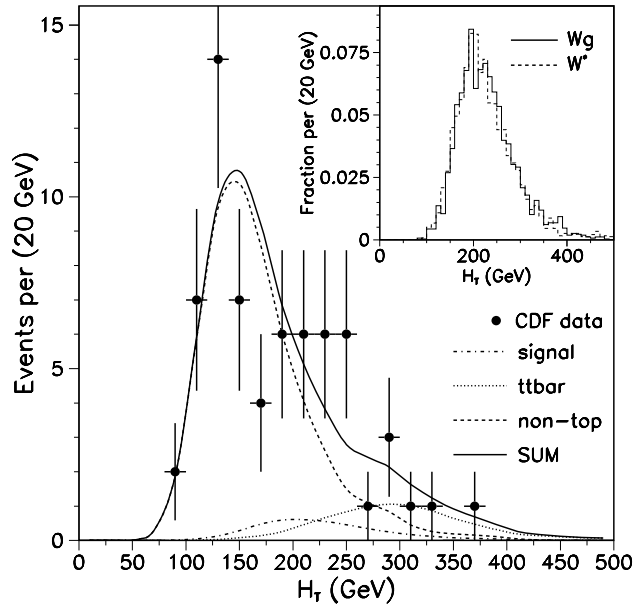,width=3.5in,height=3.5in}
}
\caption{\label{fig:ht_data_mc}
The $H_{T}$ distribution for data in the combined search, compared with 
smoothed Monte Carlo predictions for signal and backgrounds (second column in Table 
\protect\ref{tab:pred}).  $H_{T}$ is the scalar sum of $\protect\met$ and 
the transverse energies of the lepton and all jets in the event.  The inset 
shows that the Monte Carlo modeling of $H_{T}$ is very similar for both signal 
processes.}
\end{figure}

To measure the combined $Wg$ + $W^{\star}$ single-top production
cross section, we use a kinematic variable whose distribution is very similar
for the two single-top processes and is different for background processes:  
the scalar sum $H_{T}$ of $\met$ and the transverse energies of the lepton 
and all jets in the event.  We perform an unbinned maximum-likelihood fit of 
the $H_{T}$ distribution from data to a linear superposition of the expected 
$H_{T}$ distributions from single-top signal,
$t\bar{t}$ and non-top backgrounds.  We model the shape of the $H_{T}$
distribution for all sources of non-top background with {\sc vecbos}-generated 
\cite{bere91} events containing a $W$ plus two partons that we
force to be a $b\bar{b}$ pair.  We have checked that {\sc vecbos}  
reproduces the $H_{T}$ and $M_{\ell\nu b}$ distributions for the $b$-tagged 
$W+1$-jet data before the $M_{\ell\nu b}$ cut, a sample in which the  non-top 
backgrounds are expected to dominate.  In the search sample, the observed 
$H_{T}$ distribution agrees with the spectrum derived from Monte Carlo 
calculations when the latter are normalized to the a priori predicted 
numbers of events (Figure \ref{fig:ht_data_mc}).  

We set an upper limit on the cross section using the likelihood function:
\begin{displaymath}
{\cal L}(\beta_{s},\beta_{t\bar{t}},\beta_{nt}) = 
G_{1}(\beta_{t\bar{t}}) \times G_{2}(\beta_{nt})
\times {\cal L}_{\rm shape}(\beta_{s},\beta_{t\bar{t}},\beta_{nt}),
\end{displaymath}
where $\beta_{s}$, $\beta_{t\bar{t}}$ and $\beta_{nt}$ are fit parameters
representing, respectively, factors by which the standard model cross section
predictions for single-top, $t\bar{t}$ and non-top must be multiplied to fit 
the data.  The functions $G_{1}$ and $G_{2}$ are Gaussian densities
constraining the background factors $\beta_{t\bar{t}}$ and 
$\beta_{nt}$ to unity, and ${\cal L}_{\rm shape}$ represents the joint 
probability density for observing the $N_{\rm obs}$ data events at their 
respective values of $H_{T}$:
\begin{eqnarray*}
\lefteqn{{\cal L}_{\rm shape}(\beta_{s},\beta_{t\bar{t}},\beta_{nt}) = 
\frac{\mu_{\rm fit}^{N_{\rm obs}}e^{-\mu_{\rm fit}}}{N_{\rm obs}!}} \\
& &  \times \prod_{i=1}^{N_{\rm obs}}\frac{\beta_s F_s(H_{Ti}) + 
\beta_{t\bar{t}} F_{t\bar{t}}(H_{Ti}) + \beta_{nt} F_{nt}(H_{Ti})}{\mu_{\rm fit}}.
\end{eqnarray*}
In this expression, $\mu_{\rm fit}\equiv
\beta_{s}\mu_{s}+\beta_{t\bar{t}}\mu_{t\bar{t}}+\beta_{nt}\mu_{nt}$,
where $\mu_{s}$, $\mu_{t\bar{t}}$ and $\mu_{nt}$ are the predicted numbers 
of events, and $F(H_T)$ are smoothed $H_{T}$ distributions for signal and 
background, normalized to unity.  The maximum of ${\cal L}$ is obtained for 
$\beta_{s} = 2.0\pm 1.8$, where the uncertainty is statistical only and 
includes the effect of correlations with the other fit parameters.

To extract Bayesian upper limits on the single-top production rate, we 
construct a probability distribution $f(\beta_{s})$ by maximizing 
${\cal L}(\beta_{s},\beta_{t\bar{t}},\beta_{nt})$ with respect to 
$\beta_{t\bar{t}}$ and $\beta_{nt}$ for each value of $\beta_{s}$,
and multiplying the result with a flat prior distribution for $\beta_{s}$. 
We then convolute $f(\beta_{s})$ with two Gaussian smearing
functions.  The first one has width $\beta_{s}\delta n$, where
$\delta n$ is the sum in quadrature of all the normalization
uncertainties listed in Table \ref{tab:sysinc}.  The width of the second
smearing Gaussian is the sum in quadrature of all the
systematic uncertainties relative to the shape of the $H_{T}$ distribution
($\Delta S$ in Table \ref{tab:sysinc}).
Finally, the smeared distribution is integrated to find the 95\% C.L.
upper limit on single-top production.  We find this limit to be
$\beta_{s}^{.95} = 5.9$, corresponding to a cross section of 14 pb.
\begin{table}
\begin{ruledtabular}
\begin{tabular}{lcccccc}
       & \multicolumn{2}{c}{$Wg+W^{\star}$} & \multicolumn{2}{c}{$Wg$} & 
\multicolumn{2}{c}{$W^{\star}$} \\
Source & $\delta n$ & $\Delta S$ & $\delta n$ & $\Delta S$ & $\delta n$ & $\Delta S$ \\ 
\colrule
Jet $E_{T}$ scale       & 0.01 & 0.25 & 0.01 & 0.02 & 0.01 & 0.06 \\
Initial-state radiation & 0.02 & 0.15 & 0.06 & 0.07 & 0.06 & 0.13 \\
Final-state radiation   & 0.03 & 0.02 & 0.07 & 0.02 & 0.05 & 0.01 \\
Parton distributions    & 0.04 & 0.02 & 0.01 & 0.03 & 0.01 & 0.02 \\
Signal generator        & 0.02 & 0.25 & 0.08 & 0.03 & 0.07 & 0.12 \\
Background model        & -    & 0.04 & -    & 0.12 & -    & 0.18 \\
Top mass                & 0.04 & 0.01 & 0.01 & 0.12 & 0.00 & 0.35 \\
Trigger \& lepton id.   & 0.10 & -    & 0.10 & -    & 0.10 & -    \\
$b$-tag efficiency      & 0.10 & -    & 0.10 & -    & 0.10 & -    \\
Luminosity              & 0.04 & -    & 0.04 & -    & 0.04 & -    \\
\colrule
Total                   & 0.16 & 0.39 & 0.19 & 0.19 & 0.18 & 0.44 \\
\end{tabular}
\end{ruledtabular}
\caption{\label{tab:sysinc}
Systematic uncertainties on the fit result for $\beta_{s}$ in the combined
search ($Wg+W^{\star}$), and for $\beta_{Wg}$ and $\beta_{W^{\star}}$ in
the separate search (see text).  The $\delta n$ columns list fractional 
uncertainties due to signal normalization effects and the $\Delta S$ columns
absolute uncertainties due to effects on the shapes of the fitted distributions.}
\end{table}

Because of significant differences in the final-state kinematics of the two 
single-top processes, it is possible to search for them separately.  This is 
interesting, because an exotic single-top production mechanism may contribute 
to one and not the other, for example a heavy $W^{\prime}$ decaying to a 
$t\bar{b}$ quark pair adding to the apparent $W^{\star}$ rate 
\cite{rosn90}. 
For the separated search, we use events in the $W$+2-jets sample only and 
consider two non-overlapping subsamples.  The first one consists of single-tag 
events in which the reconstructed top mass lies in the window 
$145 < M_{\ell\nu b} < 205$ GeV/$c^{2}$, and the second consists of double-tag 
events. The expected compositions, calculated in the same way as for the 
combined analysis, are shown in the last two columns of Table \ref{tab:pred}: 
in the single-tag sample, $Wg$ is about 2.5 times larger than  $W^{\star}$; 
in the double-tag sample, $W^{\star}$ is about 7.5 times larger than $Wg$.

\begin{figure}
\centerline{
\epsfig{file=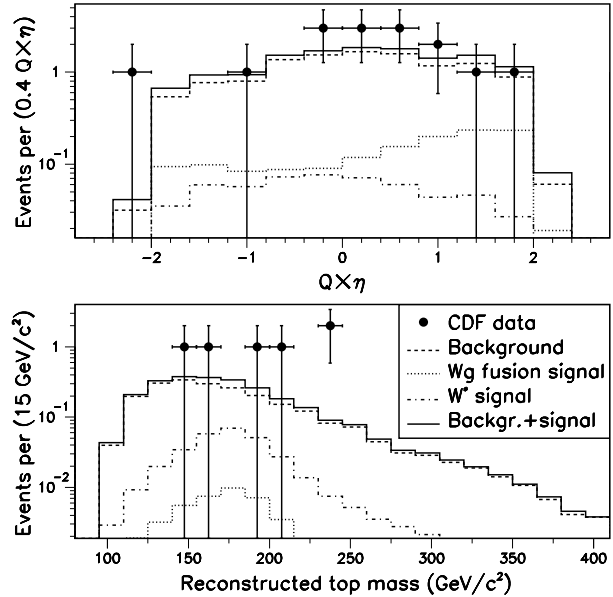,width=3.5in,height=3.5in}
}
\caption{\label{fig:qeta_fit}
Top: distribution of the product $Q\times\eta$ of the lepton charge 
and the untagged jet pseudorapidity for single-tag $W$+2-jets events.
Bottom: distribution of the reconstructed top mass for double-tag events. 
The data are compared with expectations for signal and backgrounds (third 
and fourth columns in Table \protect\ref{tab:pred}).}
\end{figure}

The $Wg$ component in the single-tag sample can be measured 
by considering that the light-quark jet in $Wg$ events
is about twice as likely to be in the same hemisphere as the outgoing
(anti)proton beam when a (anti)top quark is produced.  Thus the product
$Q\times\eta$ of the primary lepton charge and the untagged jet
pseudorapidity has a strongly asymmetric distribution.  In the 
double-tag sample, the $W^{\star}$ component can be extracted from the
distribution of $M_{\ell\nu b}$.
In this case, since both jets are tagged, the $b$-jet with the largest 
$\eta$ ($-\eta$) is used in forming $M_{\ell\nu b}$ for a $t$ ($\bar{t}$)
decay, as determined by the sign of the primary lepton in the event, 
an assignment that is expected to be correct 
64\% of the time.  The $Q \times \eta$ and $M_{\ell\nu b}$
distributions for the data are compared to expectations for signal and 
background in Figure \ref{fig:qeta_fit}. For the separate $Wg$ and $W^{\star}$
searches, we use a {\sc herwig} Monte Carlo calculation to model our signals.

A binned maximum-likelihood fit is used to
extract the amounts of $Wg$ and $W^{\star}$ present
in the $W$+2-jets data.  The likelihood function has the following form:
\begin{eqnarray*}
\lefteqn{{\cal L}(\beta_{Wg},\beta_{W^{\star}},\beta_{t\bar{t}1},\beta_{t\bar{t}2},
\beta_{nt1},\beta_{nt2}) =}\\
&   & G_{1}(\beta_{t\bar{t}1})\times G_{2}(\beta_{nt1})\times 
      {\cal L}_{1}(\beta_{Wg},\beta_{W^{\star}},\beta_{t\bar{t}1},\beta_{nt1}) 
      \times \\
&   & G_{3}(\beta_{t\bar{t}2})\times G_{4}(\beta_{nt2})\times 
      {\cal L}_{2}(\beta_{Wg},\beta_{W^{\star}},\beta_{t\bar{t}2},\beta_{nt2}),
\end{eqnarray*}
where the fit parameters are factors by which the predicted numbers of
$Wg$ ($\beta_{Wg}$), $W^{\star}$ ($\beta_{W^{\star}}$), single-tag
$t\bar{t}$ ($\beta_{t\bar{t}1}$), double-tag $t\bar{t}$ ($\beta_{t\bar{t}2}$),
single-tag non-top ($\beta_{nt1}$) and double-tag non-top ($\beta_{nt2}$)
events must be multiplied to fit the data.
The $G_i$ functions are Gaussian constraints on the normalizations
of the various backgrounds, ${\cal L}_{1}$ is a binned Poisson likelihood
for the $Q\times\eta$ distribution of single-tag events, and 
${\cal L}_{2}$ is a binned Poisson likelihood for the $M_{\ell\nu b}$
distribution of double-tag events.

The result of the maximum-likelihood fit for the single-top content of the
data is $-0.6^{+4.8}_{-4.0}$ $Wg$ events and $7.6^{+5.9}_{-4.8}$
$W^{\star}$ events.  The systematic uncertainties are listed in Table 
\ref{tab:sysinc}.  We extract upper limits on the individual single-top processes 
in the same way as for the combined search.  At the 95\% C.L., 
we find upper limits of 13 and 18 pb on single-top production in the 
$Wg$ and $W^{\star}$ channels, respectively.  These two limits
are correlated since they are derived from the same likelihood function.

In summary, we conclude that electroweak $t \bar{b}$ production is out
of reach in the Run I CDF data set. 
At the 95\% C.L., we set an upper limit on the 
combined $Wg+ W^{\star}$  single-top cross section of 14 pb.
Separate 95\% C.L. upper limits in the $Wg$ and $W^{\star}$ 
channels are 13 and 18 pb, respectively. 

\begin{acknowledgments}
We thank the Fermilab staff and the technical staffs of the
participating institutions for their vital contributions.  This work was
supported by the U.S. Department of Energy and National Science Foundation;
the Italian Istituto Nazionale di Fisica Nucleare; the Ministry of Education,
Culture, Sports, Science, and Technology of Japan; the Natural Sciences and 
Engineering Research Council of Canada; the National Science Council of the 
Republic of China; the Swiss National Science Foundation; the A. P. Sloan 
Foundation; the Bundesministerium fuer Bildung und Forschung, Germany; the 
Korea Science and Engineering Foundation (KoSEF); the Korea Research 
Foundation; and the Comision Interministerial de Ciencia y Tecnologia, Spain.
\end{acknowledgments}


\begin{thebibliography}{99}
\bibitem{CDF-coordinates}   In the CDF coordinate system, $\theta$ is the polar angle 
                    with respect to the proton beam direction and $\phi$ is the 
                    azimuth.  The pseudorapidity $\eta$ is defined as 
                    $-\ln\tan(\theta/2)$.  The transverse momentum of a particle 
                    is $P_{T}=P\sin\theta$ and its transverse energy is 
                    $E_{T}=E\sin\theta$.
\bibitem{isolation} A lepton is ``isolated'' if the non-lepton $E_T$ in an 
                    $\eta$-$\phi$ cone of radius $0.4$ centered on the lepton is less 
                    than 10\% of the lepton's $E_T$ or $P_T$.
\bibitem{missing-ET} Missing transverse energy $\met$ is defined as the magnitude of
                    $-\sum_i E_T^i \hat n_i$, where $\hat n_i$ is a unit 
                    vector in the azimuthal plane that points from the beamline
                    to the $i$th calorimeter tower.
\bibitem{wmass}     This procedure generally yields two solutions for the 
                    neutrino $P_z$. We use the solution with the smallest 
                    absolute value. In cases where the solutions are complex
                    conjugate, we use their real part.
\end{thebibliography}
\end{document}